\documentclass[a4paper,american]{paper}
\usepackage[T1]{fontenc}
\usepackage[latin9]{inputenc}
\usepackage{amsthm}
\usepackage{amsmath}
\usepackage{amssymb}

\makeatletter

\numberwithin{equation}{section}
\numberwithin{figure}{section}
\newcommand{\lyxaddress}[1]{
\par {\raggedright #1
\vspace{1.4em}
\noindent\par}
}

\makeatother

\usepackage{babel}

\begin{document}

\title{Simplification of Boltzmann Equation on $S^{3}(1)$}

\author{Lang Xia}

\maketitle

\lyxaddress{Department of Mechanics and Engineering Science, Fudan University,
Shanghai 200433, China}
\begin{abstract}
Simple form of Boltzmann equation will be proposed after introducing
a three-dimensional closed Lie group to simplify its collision term. \end{abstract}

\section{Introduction}

Navier-Stokes equation is the first order approximation of Boltzmann
equation\cite{key-1}. Therefore, solving Boltzmann equation directly
is useful to understand mysteries of fluid phenomena in detail\cite{key-2}.
The collision term in Boltzmann equation is probably the main difficulty,
and the traditional treatments are limited in Perturbation, Variation,
BGK methods and so forth\cite{key-3}, among which BGK model is the
most successful to hydrodynamics; however, this method is still limited
in treating fluid dynamics with constant temperatures and low Mach
number\cite{key-4}.

On the other hand, Lie group and Lie algebra, original in analyzing
Partial Differential Equations from the point of mathematics, is also
used to deal with Boltzmann equation. However, the corresponding solutions
only exist locally, and some of them are sensitive to the symmetrical
structures as shown in\cite{key-5,key-6,key-7,key-8,key-9}

To overcome such difficulties, a three-dimensional closed Lie group
$S^{3}(1)$ imbedded in $\mathbb{R}^{4}$, on which the collision
term can be dismissed, is introduced in this paper, and the global
solution of Boltzmann equation is discussed.

\section{Analysis of Collision Term and Results}

Two molecules are $m_{1}$ and $m_{2}$ in mass, $d_{1}$ and $d_{2}$
in diameter, $\mathbf{v}_{1}$ and $\mathbf{v}_{2}$ in velocity before
collision, and $\mathbf{w}_{1}$ and $\mathbf{w}_{2}$ after collision,
respectively. The collision between molecules is imperfect elastic.
In terms of momentum theorem and energy conservation law, we have

\begin{equation}
\begin{cases}
m_{1}\mathbf{v}_{1}+m_{2}\mathbf{v}_{2}=m_{1}\mathbf{w}_{1}+m_{2}\mathbf{w}_{2}\\
\frac{1}{2}m_{1}\mathbf{v}_{1}^{2}+\frac{1}{2}m_{2}\mathbf{v}_{2}^{2}=\frac{1}{2}m_{1}\mathbf{w}_{1}^{2}+\frac{1}{2}m_{2}\mathbf{w}_{2}^{2}+\triangle E\end{cases}\label{eq:1}\end{equation}
where \begin{equation}
\triangle E=\frac{1}{2}(1-\epsilon^{2})\frac{m_{1}m_{2}}{m_{1}+m_{2}}(v_{1}-v_{2})^{2}\label{eq:2}\end{equation}
 and $\epsilon$ is the restitution coefficient. Define $\mid\mathbf{d}\mid$
the distance between centers of $m_{1}$and $m_{2}$. Let $\mathbf{n}=\frac{\mathbf{d}}{\mid\mathbf{d}\mid}$
and

\begin{equation}
\begin{cases}
\mathbf{w}_{1}-\mathbf{v}_{1}=\lambda_{1}\mathbf{n}\\
\mathbf{w}_{2}-\mathbf{v}_{2}=\lambda_{2}\mathbf{n}\end{cases}\label{eq:3}\end{equation}

From Eq.2.1 and Eq.2.3, two cases of collision are deduced as

\begin{equation}
\begin{cases}
\mathbf{w}_{1}=\mathbf{v}_{1}+(1+\epsilon)\frac{m_{2}}{m_{1}+m_{2}}[(v_{2}-v_{1})\cdot\mathbf{n}]\mathbf{n}\\
\mathbf{w}_{2}=\mathbf{v}_{2}-(1+\epsilon)\frac{m_{1}}{m_{1}+m_{2}}[(v_{2}-v_{1})\cdot\mathbf{n}]\mathbf{n}\end{cases}\label{eq:4}\end{equation}
and 

\begin{equation}
\begin{cases}
\mathbf{w}_{1}=\mathbf{v}_{1}+(1-\epsilon)\frac{m_{2}}{m_{1}+m_{2}}[(v_{2}-v_{1})\cdot\mathbf{n}]\mathbf{n}\\
\mathbf{w}_{2}=\mathbf{v}_{2}-(1-\epsilon)\frac{m_{1}}{m_{1}+m_{2}}[(v_{2}-v_{1})\cdot\mathbf{n}]\mathbf{n}\end{cases}\label{eq:5}\end{equation}
both of them may exist in experiment as shown in \cite{key-1}. Write
the collision term of Boltzmann equation as\cite{key-1}

\begin{equation}
\frac{\partial f_{1}}{\partial t}\mid_{coll}=\iint(J^{*}f_{1}^{'}f_{2}^{'}-f_{1}f_{2})\frac{1}{4}d^{2}(\mathbf{v}_{1}-\mathbf{v}_{2})\cdot\mathbf{n}d\mathbf{v}_{2}d\Omega\label{eq:6}\end{equation}
where $f(\mathbf{r},\mathbf{v},t)$ is an one-particle probability
distribution function; $f,\; f^{'}$ denote the one-particle probability
distribution function before and after collision. $\Omega$ is the
scattering angle of the binary collision $\{\mathbf{W}_{2},\mathbf{W}_{1}\}\rightarrow\{\mathbf{v}_{2},\mathbf{v}_{1}\}$.
Here $J^{*}$ is the Jacobean matrix defined as

\begin{equation}
J^{*}=\frac{\partial(\mathbf{W}_{2},\mathbf{W}_{1})}{\partial(\mathbf{v}_{2},\mathbf{v}_{1})}\label{eq:7}\end{equation}

In terms of Eq.2.4 and Eq.2.5

\begin{equation}
J^{*}=\epsilon\label{eq:8}\end{equation}

Write Eq.2.6 in general form as matter of convenience

\begin{equation}
\frac{\partial f}{\partial t}\mid_{coll}=\iint(\epsilon f^{'}f_{1}^{'}-ff_{1})\frac{1}{4}d^{2}(\mathbf{v}-\mathbf{v}_{1})\cdot\mathbf{n}d\mathbf{v}_{1}d\Omega\label{eq:9}\end{equation}

In order to treat the above collision term, we shall first introduce
a four-dimensional Euclidean space. Let $\lambda\in\mathbb{R}\setminus\{0\}$,
$(v_{1},v_{2},v_{3},v_{4})\in\mathbb{R}^{4}$, such that \begin{equation}
v_{1}^{2}+v_{2}^{2}+v_{3}^{2}+v_{4}^{2}=\lambda^{2}\label{eq:10}\end{equation}
 then the following differentiable manifold

\[
M=S^{3}(1)=\{(\vartheta_{1},\vartheta_{2},\vartheta_{3},\vartheta_{4})\in\mathbb{R}^{4}\mid\sum_{i=1}^{4}\vartheta_{i}^{2}=1\}\]
is a Lie group\cite{key-10}, where $\mathbf{\vartheta}=\frac{\mathbf{v}}{\lambda}$.

According to \cite{key-11}, Boltzmann equation can be written as

\begin{equation}
\frac{\partial f}{\partial t}+\mathbf{v}\cdot\nabla f+\frac{\mathbf{F}}{m}\cdot\nabla_{\mathbf{v}}f=\frac{\partial f}{\partial t}\mid_{coll}\label{eq:11}\end{equation}

Recall the General Stokes Formula\cite{key-12}

\begin{equation}
\int_{\partial M}F=\int_{M}dF\label{eq:12}\end{equation}
where the boundary $\partial M$ of $M$ is smooth and simple, $F\:\in C^{\infty}(M)$,
we can obtain 

\[
\frac{\partial f}{\partial t}\mid_{coll}=\int_{s}[\int_{\partial M}(\epsilon f^{'}f_{1}^{'}-ff_{1})\frac{1}{4}d^{2}(\mathbf{v}-\mathbf{v}_{1})\cdot\mathbf{n}d\mathbf{v}_{1}]d\Omega\]

\begin{equation}
=\int_{s}[\int_{M}\frac{\partial[(\epsilon f^{'}f_{1}^{'}-ff_{1})\frac{1}{4}d^{2}(\mathbf{v}-\mathbf{v}_{1})\cdot\mathbf{n}]}{\partial v_{4}}dv_{4}d\mathbf{v}_{1}]d\Omega\label{eq:13}\end{equation}

Here we can see the collision term on the Lie group $M$ . Suppose
$(\epsilon f^{'}f_{1}^{'}-ff_{1})\frac{1}{4}d^{2}(\mathbf{v}-\mathbf{v}_{1})\cdot\mathbf{n}$
to be smooth. Since it is independent to $v_{4}$, then we have

\begin{equation}
\frac{\partial f}{\partial t}\mid_{coll}=0\label{eq:14}\end{equation}

Consequently, the Boltzmann equation can be written as Vlasov-Poisson
equation\cite{key-13}

\begin{equation}
\frac{\partial f}{\partial t}+\mathbf{v}\cdot\nabla f+\frac{\mathbf{F}}{m}\cdot\nabla_{\mathbf{v}}f=0\label{eq:15}\end{equation}

Suppose $e$ the identity in $M$, let $p=(0,0,0,1)$, $U=S^{3}(1)\setminus\{p\}$.
Define $\varphi:\: U\:\longrightarrow\mathbb{R}^{3}$, thus $(U,\:\varphi)$
is a chart of $M$ containing identity $e$. where 

\begin{equation}
\varphi(v_{1},v_{2},v_{3},v_{4})=(v_{1}^{*},v_{2}^{*},v_{3}^{*})=(\frac{v_{1}}{1-v_{4}},\frac{v_{2}}{1-v_{4}},\frac{v_{3}}{1-v_{4}})\label{eq:16}\end{equation}

Here let $\mathbf{v}=\vartheta$ as a matter of convenience.

Eq.15 can be easily written in the form of

\begin{equation}
\frac{df}{dt}+\frac{\mathbf{F}}{m}\cdot\nabla_{\mathbf{v}}f=0\label{eq:17}\end{equation}
where $\frac{df}{dt}=\frac{\partial f}{\partial t}+\mathbf{v}\cdot\nabla f$,
and from Eq.2.16

\begin{equation}
\frac{\partial}{\partial v_{i}}=\frac{\partial}{\partial v_{j}^{*}}\frac{\partial v_{j}^{*}}{\partial v_{i}}\label{eq:18}\end{equation}

Let $J=\mid\frac{\partial v_{j}^{*}}{\partial v_{i}}\mid$, and write
Eq.2.15 as

\begin{equation}
\frac{df}{dt}+J\frac{\mathbf{F}}{m}\cdot\nabla_{\mathbf{v^{*}}}f=0\label{eq:19}\end{equation}

Let $G(\tau)$ be the one-parameter subgroup on Lie group $M$, and
define 

\[
X=\frac{JF_{i}}{m}\frac{\partial}{\partial v_{i}^{*}}\]
which is treated as Lie algebra on $M$; usually, we suppose constant
acceleration of molecules\cite{key-11} and the external force $\mathbf{F}$
to be independent with time $t$ such as gravitation. Recall the definition
of tangent vectors on manifold and the character of one-parameter
subgroup\cite{key-10,key-14}

\begin{equation}
dG(\frac{d}{d\tau}(0))f=\frac{d}{d\tau}(f\circ G)(0)=Xf(e)\label{eq:20}\end{equation}

Here we suppose $f\,\in C^{\infty}(e),\: f\circ G\,\in C^{\infty}(0)$
. Then Eq.2.18 can be written as

\begin{equation}
\frac{df}{dt}(e)+\frac{d}{d\tau}(f\circ G)(0)=0\label{eq:21}\end{equation}

Since M is a global Lie group, we can write the above equation in
the form of

\begin{equation}
\frac{df}{dt}+\frac{d}{d\tau}(f\circ G)=0\label{eq:22}\end{equation}

Moreover, let $\tau=\theta(t)$ , and $d\tau=\theta^{'}(t)dt$, then
integrate Eq.2.22

\begin{equation}
f(\mathbf{v},t)+\theta^{'}(t)f(G(\theta(t)))=C(\mathbf{v})\label{eq:23}\end{equation}
where $C(\mathbf{v})$ is independent with $t$, and determined by
boundary and initial conditions; $G(t)$ is a known function.

\section{Conclusion}

We firstly derived the collision term in a general process with restitution
coefficient. However, this coefficient makes no difference to ours
analysis, for the collision term in Boltzmann equation is bound to
disappear as long as it is on a closed differentiable manifold. At
the same time, we can always introduce a higher dimensional space
for any $\mathbf{v}=(v_{1},v_{2},v_{3})\in\mathbb{R}^{3}$ , such
that 

\[
v_{1}^{2}+v_{2}^{2}+v_{3}^{2}+v_{4}^{2}=\lambda^{2}\]
where the parameter $\lambda$ controls the solution of Boltzmann
equation, and manifold $S^{3}(1)$ is a global Lie group. So the differentiable
manifold is well-defined and the corresponding results are global.

\section{Acknowledgments}

The author thanks Dr. Heng Ren for his kind help in collecting literature,
particualrly thanks Prof. Mingqing You.

\end{document}